\documentclass[aps,onecolumn,superscriptaddress,prl,floatfix,british]{revtex4}
\usepackage{amssymb}
\usepackage{amsmath}
\usepackage{babel}
\usepackage[sort&compress]{natbib}
\usepackage{topcapt}
\usepackage{multirow}

\usepackage{graphicx}
\usepackage{subfigure}
\usepackage{multirow}
\usepackage{diagbox} 
\usepackage{booktabs} 
\usepackage{ulem}

\usepackage[colorlinks,linkcolor=dred,urlcolor=dblue,citecolor=dgreen]{hyperref}
\usepackage{stmaryrd}
\usepackage{amsfonts}
\usepackage{mathrsfs}
\usepackage{txfonts}
\graphicspath{{figs/}}

\usepackage{xcolor}
\usepackage{tikz}
\usepackage{circuitikz}
\usetikzlibrary{backgrounds,fit,decorations.pathreplacing,calc,backgrounds}  
\usepackage{booktabs}

\usepackage{color}
\usepackage{pdfpages}
\definecolor{dred}{rgb}{.8,0.2,.2}
\definecolor{ddred}{rgb}{.8,0.5,.5}
\definecolor{dblue}{rgb}{.2,0.2,.8}
\definecolor{dgreen}{rgb}{.2,0.5,.2}

\usepackage{multirow}

\newcommand{\ket}[1]{| #1 \rangle}

\newcolumntype{I}{!{\vrule width 1pt}}

\def\be{\begin{eqnarray}}
\def\ee{\end{eqnarray}}

\usepackage{times}

\definecolor{Pr}{rgb}{0.4,0.3,0.9}

\begin{document}
\title{A Quantum Convolutional Neural Network on NISQ Devices}
\author{ShiJie Wei}
\email{weisj@baqis.ac.cn}
\affiliation{Beijing Academy of Quantum Information Sciences, Beijing 100193, China}
\affiliation{State Key Laboratory of Low-Dimensional Quantum Physics and Department of Physics, Tsinghua University, Beijing 100084, China}
\author{YanHu Chen}
\affiliation{Institute of Information Photonics and Optical Communications, Beijing University of Posts and Telecommunications, Beijing 100876, China}
\author{ZengRong Zhou}
\affiliation{Beijing Academy of Quantum Information Sciences,  Beijing 100193, China}
\affiliation{State Key Laboratory of Low-Dimensional Quantum Physics and Department of Physics, Tsinghua University, Beijing 100084, China}
\author{GuiLu Long}
\email{gllong@tsinghua.edu.cn}
\affiliation{State Key Laboratory of Low-Dimensional Quantum Physics and Department of Physics, Tsinghua University, Beijing 100084, China}
\affiliation{Beijing Academy of Quantum Information Sciences,  Beijing 100193, China}
\affiliation{ Beijing National Research Center for Information Science and Technology and School of Information Tsinghua University, Beijing 100084, China}
\affiliation{Frontier Science Center for Quantum Information, Beijing 100084, China}
\begin{abstract}
  Quantum machine learning is one of the most promising applications of quantum computing in the Noisy Intermediate-Scale Quantum(NISQ) era. 
  Here we propose a quantum convolutional neural network(QCNN) inspired by convolutional neural networks(CNN), which greatly reduces the computing complexity compared with its classical counterparts,  with  $O((log_{2}M)^6) $ basic gates and $O(m^2+e)$ variational parameters,  where $M$ is the input data size, $m$ is the filter mask size and  $e$ is the number of parameters in a Hamiltonian. Our model is robust to certain noise for image recognition tasks and the parameters are independent on the input sizes, making it friendly to near-term quantum devices. We demonstrate QCNN with two explicit examples
  . First, QCNN is applied to image processing and   numerical simulation of three types of spatial filtering, image smoothing, sharpening, and edge detection are performed. Secondly, we demonstrate QCNN in recognizing image, namely, the recognition of handwritten numbers. Compared with previous work, this machine learning model can provide implementable quantum circuits that  accurately  corresponds to a specific classical  convolutional kernel. It provides an efficient avenue to transform CNN to QCNN directly and opens up the prospect of exploiting quantum power to process information in the era of big data.
\end{abstract}

\maketitle

\section{Introduction}\label{Intro}
Machine learning has fundamentally transformed the way people think and behave. Convolutional Neural Network(CNN) is an important machine learning model which has the advantage of utilizing the correlation information of data, with many interesting applications ranging from image recognition to precision medicine. 

 Quantum information processing (QIP)\cite{benioff1980computer,feynman1982simulating}, which exploits  quantum-mechanical phenomena such as quantum superpositions and quantum entanglement, allows one to overcome the limitations of classical computation and reaches higher computational speed for certain problems\cite{ShorFactor,grover1997quantum, Grover_Long_2001}.
Quantum machine learning, as an interdisciplinary study between machine learning and quantum information, has undergone a flurry of developments in recent years\cite{biamonte2017quantum,dunjko2016quantum,killoran2019continuous,liu2019hybrid,hu2019quantum, farhi2020classification,huggins2019towards,yuan2020quantum,zhang2020recent,liu2021solving}.  Machine learning algorithm consists of three components: representation, evaluation and optimization, and  the quantum version\cite{farhi2018classification,cong2019quantum,britt2020modeling,schuld2019quantum,li2020quantum} usually concentrates on realizing the evaluation part, the  fundamental construct in deep learning\cite{goodfellow2016deep}. 

A CNN generally consists of three layers, convolution layers, pooling layers and fully connected layers. The convolution layer calculates new pixel values $x_{ij}^{(\ell)}$ from a linear combination of the neighborhood pixels in the preceding map with the specific weights, $x_{i,j}^{(\ell)} = \sum_{a,b=1}^m w_{a,b} x_{i+a-2,j+b-2}^{(\ell -1)}$, where the weights $w_{a,b}$ form a $m \times m$ matrix named as a convolution kernel or a filter mask. Pooling layer reduces feature map size, e.g. by taking the average value from four contiguous pixels, and are often followed by application of a nonlinear (activation) function. The fully connected layer computes the final output by a linear combination of all remaining pixels with specific weights determined by parameters in  a fully connected layer. The weights in the filter mask and fully connected layer are optimized by training on large datasets. 

In this article, we demonstrate the basic framework of a quantum convolutional neural network(QCNN) by sequentially realizing convolution layers, pooling layers and fully connected layers. Firstly, we implement convolution layers based on linear combination of unitary operators(LCU)\cite{gui2006general,gudder2007mathematical,wei2016duality}. Secondly, we abandon some qubits in the quantum circuit to simulate the effect of the classical pooling layer. Finally, the fully connected layer  is realized by measuring the expectation value of a parametrized  Hamiltonian and then a nonlinear (activation) function to post-process the expectation value.
We perform numerical  demonstrations with two examples to show the validity of our  algorithm. Finally, the  computing complexity of our algorithm is discussed followed by a summary.

\section{Framework of Quantum Neural Networks}\label{sec:framework}
\subsection{Quantum Convolution Layer}
\begin{figure}[htb]
\centerline{
\includegraphics[scale=0.5]{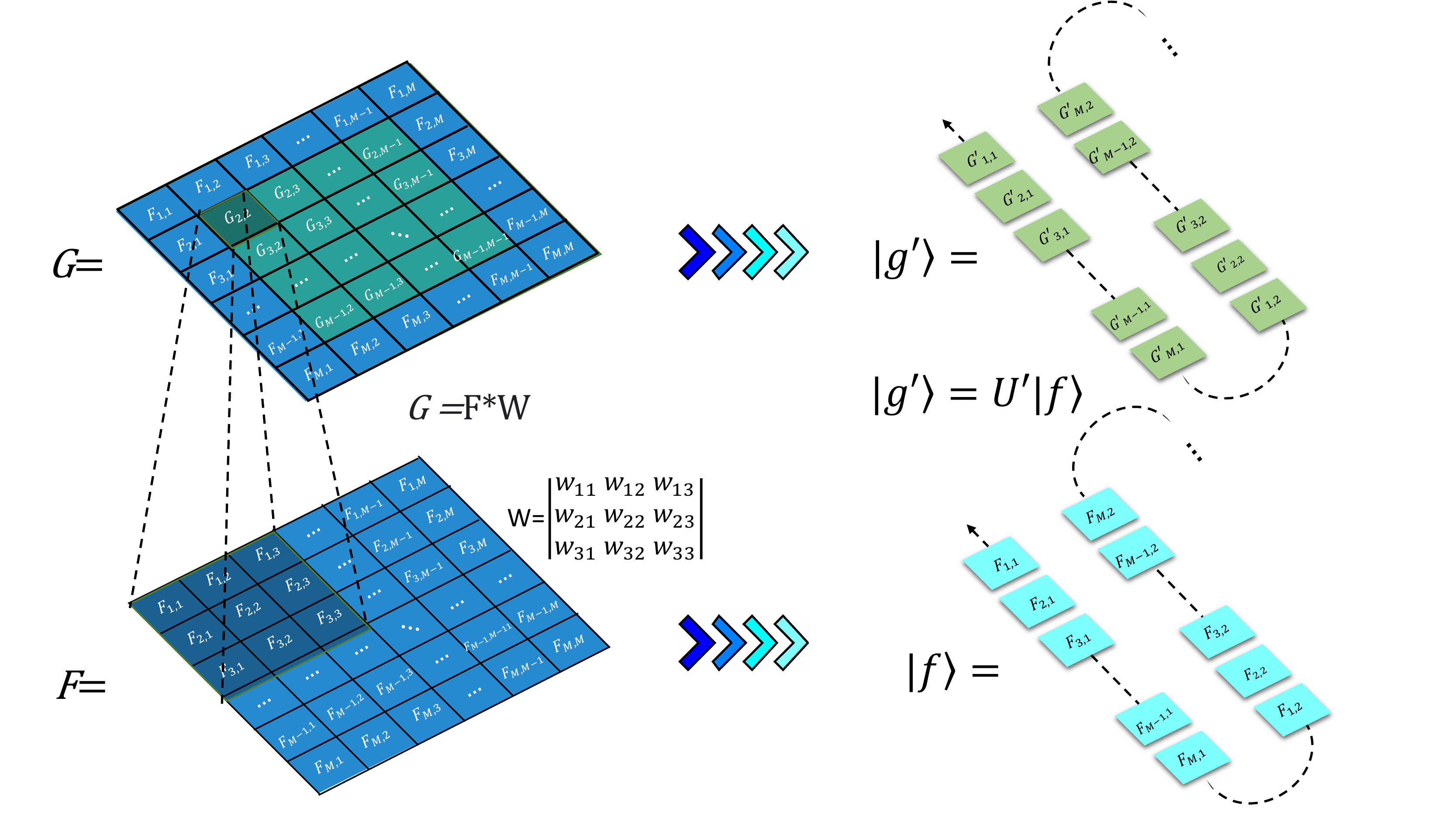}
}
\caption{Comparison of classical  convolution processing and quantum  convolution processing. $F$ and $G$ are the input and output image  data, respectively. On the classical computer, a $M \times M$ image can be represented as a matrix and encoded with at least $2^n$ bits [$n=\lceil \log_2 (M^2)\rceil$]. The classical image transformation through the convolution layer  is performed by matrix computation $F*W$, which leads to $x_{i,j}^{(\ell)} = \sum_{a,b=1}^m w_{a,b} x_{i+a-2,j+b-2}^{(\ell -1)}$.  The same image can be represented as a quantum state and encoded in at least $n$ qubits on a quantum computer. The quantum image transformation is realized by the unitary evolution $U$ on a specific quantum state.}
\label{fig:Quan Cov}
\end{figure}

The first step for performing quantum convolution layer is to encode  the  image data into a quantum system. In this work, we  encode  the pixel positions in the computational basis states and the pixel values in the probability amplitudes, forming a pure quantum state. 
Given a 2D image $F=(F_{i,j})_{M \times L}$, where $F_{i,j}$ represents the pixel value at position $(i,j)$ with $i = 1,\dots, M$ and $j = 1,\dots,L$.  $F$ is transformed as a vector $\vec{f}$ with $ML$ elements  by puting the  first column of $F$ into the first $M$ elements of $\vec{f}$, the second column the next $M$ elements , etc. That is,
 \begin{eqnarray}
\vec{f}=(F_{1,1},F_{2,1},\dots,F_{M,1},F_{1,2},\dots,F_{i,j},\dots,F_{M,L})^T.
\end{eqnarray}
Accordingly, the image data $\vec{f}$ can be mapped onto a pure quantum state $\ket{f} = \sum_{k=0}^{2^{n}-1} c_k \ket{k}$ with $ n=\lceil \log_2 (ML)\rceil$ qubits, where the computational basis $ \ket{k}$ encodes the position $(i,j)$ of each pixel, and the coefficient $c_k$ encodes the pixel value, i.e., $c_k = F_{i,j}/(\sum {F_{i,j}^2})^{1/2}$ for $k < ML$ and $c_k = 0 $ for $k \geq ML$. Here $(\sum {F_{i,j}^2})^{1/2}$ is a constant factor to normalizing the quantum state.

Without loss of generality, we focus on  the input image with $M=L=2^n$ pixels. The convolution layer transforms an input image $F=(F_{i,j})_{M \times M}$ into an output image $G=(G_{i,j})_{M \times M}$ by  a specific filter mask $W$. In the quantum context, this linear transformation, corresponding to a specific spatial  filter operation, can be represented as $\ket{g}=U \ket{f}$ with the input image state $\ket{f}$ and the output image state $\ket{g}$.  For simplicity, we take a  $3\times 3$ filter mask as an example
\begin{equation}\label{mask}
  W=\left[\begin{matrix}
  w_{11} & w_{12} & w_{13}\\
  w_{21} & w_{22} & w_{23}\\
  w_{31} & w_{32} & w_{33}\\
  \end{matrix}\right].
  \end{equation}
The generalization to arbitrary $m\times m$ filter mask is straightforward.
Convolution operation will transform the input image $F=(F_{i,j})_{M \times M}$ into the output image as $G=(G_{i,j})_{M \times M}$ with the pixel $G_{i,j} = \sum_{u,v =1}^{3} w_{uv} F_{i+u-2,j+v-2}$ ($2\leq i, j \leq M-1$). The corresponding quantum evolution $U\ket{f}$ can be performed as follows. We represent input image $F=(F_{i,j})_{M \times M}$  as an initial state 
 \begin{eqnarray}
  \ket{f} = \sum_{k=0}^{M^2-1} c_k \ket{k},
\end{eqnarray}
where $c_k = F_{i,j}/(\sum {F_{i,j}^2})^{1/2}$. The $ M^{2} \times M^{2}$ linear filtering operator $ U $ can be defined as\cite{yao2017quantum}: 
 \begin{equation}\label{evolution}
  U=\left[\begin{matrix}
 E &  & &   &  &      \\
 V_1 & V_2 & V_3  &      & &   \\
 & \ddots & \ddots & \ddots  & &\\
 &  & \ddots & \ddots &\ddots  &\\
 & & & V_1 & V_2 & V_3\\
   & &  & &  & E    \\
\end{matrix}\right],
\end{equation}
where $E$ is an $M$ dimensional identity matrix, and $V_1$, $V_2$, $V_3$ are $M\times M$ matrices defined by

\begin{equation}
	V_1=\left(                 
	\begin{array}{ccccc}  
	0 &  & & &         \\
	w_{11} & w_{21} & w_{31}  &  &   \\
	& \ddots & \ddots & \ddots & \\
	&  &  w_{11} &w_{21} & w_{31}\\
	& &  & &  0     \\
	\end{array}
	\right)_{M\times M}
	V_2=\left(                 
	\begin{array}{ccccc}  
		1 &  & &   &     \\
		w_{12} & w_{22} & w_{32}  &  &  \\
		& \ddots & \ddots & \ddots& \\

		&  & w_{12} &w_{22} & w_{32}\\
		& & &  &  1     \\
	\end{array}
	\right)_{M\times M}  
	V_3=\left(
	\begin{array}{ccccc}  
	0 &  & &   &       \\
	w_{13} & w_{23} & w_{33}  &    &     \\
	& \ddots & \ddots & \ddots  & \\
	& &   w_{13} &w_{23} & w_{33}\\
	& &   &  & 0     \\
	\end{array}
	\right)_{M\times M}.
\end{equation}
 Generally speaking, the linear filtering operator $ U$  is non-unitary that can not be performed directly. Actually, we can embed $ U $ in a bigger system with an ancillary system and decompose it into a linear combination of four unitary operators \cite{xin2020quantum}. 
$U=U_1+U_2+U_3+U_4$, where $U_1=(U+U^{\dagger})/2 +i \sqrt{I-(U+U^{\dagger})^2/4}$, $U_2=(U+U^{\dagger})/2 -i \sqrt{I-(U+U^{\dagger})^2/4}$,  $U_3=(U-U^{\dagger})/2i +i \sqrt{I+(U-U^{\dagger})^2/4}$  and $U_4=(U-U^{\dagger})/2i -i \sqrt{I+(U-U^{\dagger})^2/4}$.  
However, the basic gates consumed to perform $U_i$ scale exponentially in the dimensions of quantum systems, making the quantum advantage diminishing. Therefore, we present a new approach to construct the filter operator to reduce the gate complexity. For convenience, we change the elements of the first row, the last row, the first  column  and  the last column in the matrix $V_1, V_2, $ and $ V_3$, which is allowable in  imagining processing, to the following form

\begin{equation}
	V'_1=\left(                 
	\begin{array}{ccccc}  
    w_{21} & w_{31} & &   &  w_{11}     \\
	w_{11} & w_{21} & w_{31}  &  &   \\
	& \ddots & \ddots & \ddots & \\
	&  &  w_{11} &w_{21} & w_{31}\\
  w_{31} & &   &  w_{11} & w_{21}    \\
	\end{array}
	\right)_{M\times M}
	V'_2=\left(                 
	\begin{array}{ccccc}  
		w_{22}  & w_{32}  & &  &      w_{12}    \\
		w_{12} & w_{22} & w_{32}  &  &  \\
		& \ddots & \ddots & \ddots& \\

		&  & w_{12} &w_{22} & w_{32}\\
    w_{32} & &  & w_{12}  & w_{22}     \\
	\end{array}
	\right)_{M\times M} 
	V'_3=\left(
	\begin{array}{ccccc}  
    w_{23}  &  w_{33} & &     & w_{13}      \\
	w_{13} & w_{23} & w_{33}  &    &     \\
	& \ddots & \ddots & \ddots  & \\
	& &   w_{13} &w_{23} & w_{33}\\
	w_{33}   &  & & w_{13}  & w_{23}      \\
	\end{array}
	\right)_{M\times M}.
\end{equation}

Defining the adjusted linear filtering operator $ U' $ as 
\begin{equation}\label{evo2}
  U' = \left[\begin{matrix}
 V'_2 & V'_3 & &  &   &    V'_1   \\
 V'_1 & V'_2 & V'_3  &      &  &   \\
 & \ddots & \ddots & \ddots & &\\
 &  & \ddots & \ddots &\ddots  &\\
 & & & V'_1 & V'_2 & V'_3\\
  V'_3 & &  & & V'_1  & V'_2   \\
\end{matrix}\right],
\end{equation}

 Next, we decompose $V'_{\mu}(\mu=1,2,3)$ into three unitary matrices without normalization, $V'_{\mu}=V'_{1 \mu}+V'_{2 \mu}+V'_{3 \mu}$, where

 \begin{equation}
V'_{1 \mu}=\left(
\begin{array}{ccccc}  
 &  & &   &  w_{1\mu}     \\
w_{1\mu} &     &    &  &   \\
\ddots & \ddots & \ddots &  &\\
& &  w_{1\mu} & & \\
 & &   & w_{1\mu} &   \\
\end{array}\right)_{M\times M}\\
V'_{2 \mu}=\left(
\begin{array}{ccccc}  
w_{2\mu}  &   &   &  &      \\
 & w_{2\mu} &      &  &   \\
 &   \ddots & \ddots &\ddots & \\
 & &   &w_{2\mu} & \\
& &  &  & w_{2\mu}     \\
\end{array}\right)_{M\times M}\\
V'_{3\mu}=\left(
\begin{array}{ccccc}  
 &  w_{3\mu} & &     &     \\
 &  & w_{3\mu}  &      &   \\
 &  &\ddots \ddots & \ddots & \\
 & &   & & w_{3\mu}\\
 w_{3\mu}   & &   &   &   \\
\end{array}\right)_{M\times M}.
\end{equation}

 Thus, the linear filtering operator $ U'$ can be expressed as
\begin{eqnarray}
U'=\sum_{\mu=1}^{3} \sum_{v=1}^{3}(V'_{\mu\mu}/w_{\mu\mu})\otimes V'_{v\mu}.
\end{eqnarray}

which can be  simplified to
\begin{eqnarray}
  U'=\sum_{k=1}^{9}\beta_{k} Q_{k},
\end{eqnarray}
where $ Q_{k}=(V'_{\mu\mu}/w_{\mu\mu})\otimes V'_{v\mu}/w_{v\mu} $ is unitary, and $\beta_{k}$ is a relabelling of the indices.

\begin{figure}
\centering
\includegraphics[width=0.9\textwidth]{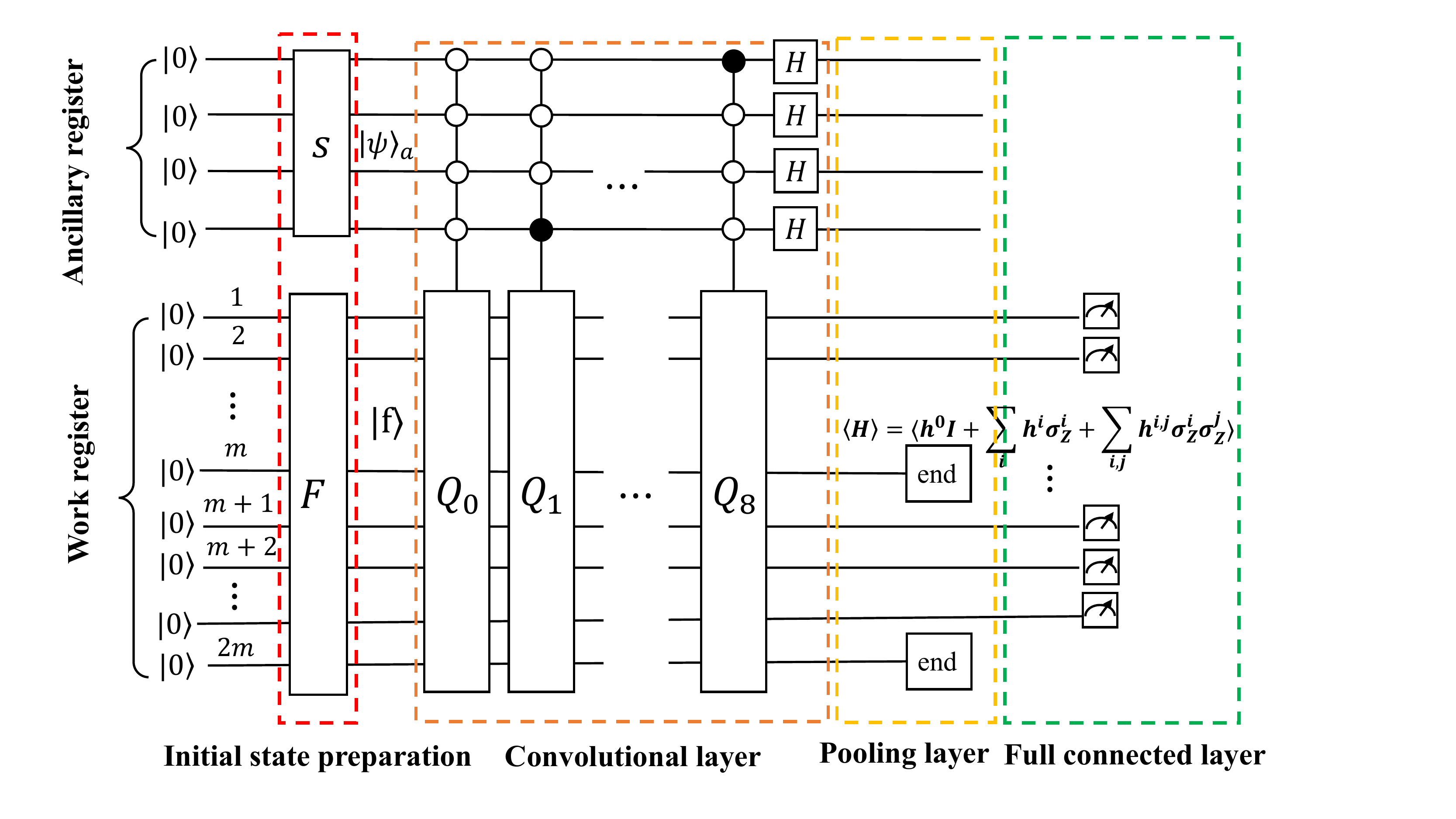}
\caption{Quantum circuit for realizing the QCNN. $|f\rangle$ denotes the initial state of work system after encoded the image data, and the ancillary system is a four qubits system in the $|0000\rangle_a$ state. The squares represent unitary operations and the circles represent the state of the controlling system. Unitary operations $Q_1$,$Q_2$,$\cdots$, $Q_9$, are activated only when the auxiliary system is in state $|0000\rangle$,$|0001\rangle$$\cdots$,$|1000\rangle$  respectively.} \label{im2}
\end{figure}

Now, we can perform $ U'$ through the linear combination of unitary operators $Q_{k}$. The number of unitary operators is equal to the size of filter mask. The quantum circuit to realize $ U' $ is shown in Fig.{\ref{im2}}. The work register $\ket{f}$ and four ancillary qubits $\ket{0000}_a$  are entangled together to form a bigger system.

Firstly, we  prepare the  initial state  $\ket{f}$ using amplitude encoding method or quantum random access memory(qRAM). Then, performing unitary matrix $S$ on the  ancillary registers to transform $ |0000\rangle_a $  into a specific superposition state $|\psi\rangle_a$
\begin{eqnarray}\label{Superstate}
S|0000\rangle_a=|\psi\rangle_a= \sum_{v=1}^{9}\beta_{k}/N|k\rangle
\end{eqnarray}
where $N_c=\sqrt{\sum_{k=1}^{9}\beta_{k}^2}$ and $ S$ satisfies
\begin{equation}
  S_{k,1} = \begin{cases}\beta_{k}/N_c   &\textrm{if $k\leq 9$}\\
 0  & \textrm{if $k > 9$.}\end{cases}\label{eq:s}
\end{equation}
$ S $ is a parameter matrix corresponding to a specific filter mask that realizes a specific task.

Then, we implement a series of ancillary system controlled operations $Q_{k} \otimes |k\rangle \langle k|$ on the work system $\ket{f}$ to realize  LCU.
Nextly,  Hadamard gates $H^T=H^{\otimes4}$ are acted to uncompute the ancillary registers $|\psi\rangle_a$. The state is transformed to
\begin{eqnarray}
  \ket{g'}= \sum_{i=1}^{16}\frac{1}{N_c}|i\rangle 
    \sum_{k=1}^{16} H^T_{(i:)}S_{(:1)} Q_{k}\ket{f},
      \label{eqH}
    \end{eqnarray}
    where $H^T_{(i:)}$ is the i-th row in matrix $H^T$ and $S_{(:1)}$ is the first column in matrix $S$.
The first term  equals to 
\begin{eqnarray}
\frac{1}{N_c}|0\rangle 
\sum_{k=1}^{9} \beta_{k} Q_{k}\ket{f},
  \label{eq3}
\end{eqnarray}
which corresponds to the filter mask $W$. The $i$-th term equals to filter mask $W^i(i=2,3,\dots,16)$, where    

\begin{equation}\label{mask2}
    W^i=\left[\begin{matrix}
    H^T_{i1}w_{11} & H^T_{i4}w_{12} & H^T_{i7}w_{13}\\
    H^T_{i2}w_{21} & H^T_{i5}w_{22} & H^T_{i8}w_{23}\\
    H^T_{i3}w_{31} & H^T_{i6}w_{32} & H^T_{i9}w_{33}\\
    \end{matrix}\right].
    \end{equation}
Totally, $16$ filter masks are realized, corresponding to ancilla qubits in $16$ different state $\ket{i}(i=1,2,\dots,16)$. Therefore, the whole effect of evolution on state $\ket{f}$ without considering the ancilla qubits, is the linear combination of the effects of 16 filter masks.

If we only need one filter mask $W$, measuring the ancillary register and conditioned on seeing $|0000\rangle$. We have the state $\frac{1}{N_c}|0000\rangle U'\ket{f}$, which is proportional to our expected result state $ \ket{g}$. The probability of detecting the ancillary state $|0000\rangle$ is
\begin{eqnarray}
P_{s}= \parallel\sum_{k=1}^{9}\beta_{k} Q_{k} \ket{f}\parallel^{2}/{N_c^2}   
\end{eqnarray}

After obtaining the final result $\frac{1}{N_c}U'|f\rangle $, we can multiply the constant factor $N_c $ to compute $\ket{g'}=U'|f\rangle$. In conclusion,  the filter operator $U'$ can be decomposed into a linear combination of nine unitary operators in the  case that the general filter mask is $ W $. Only four qubits or a nine energy level ancillary system is consumed to realize the general filter operator $ U'$, which is independent on the dimensions of image size.

The final stage of our method is to extract useful information from the processed results $\ket{g'}$. Clearly,  the image state $\ket{g}$ is different from $\ket{g'}$. However, not all elements in  $\ket{f}$ are evaluated, the elements corresponding to the four edges of original image remain unchanged. One is only interested in the pixel values which are evaluated by $W$ in $\ket{f}$. These pixel values in $\ket{g'}$  are as same as that in $\ket{g}$(see proof in supplementary materials). So, we can obtain the informations of $G=(G_{i,j})_{M \times M}$ ($2\leq i, j \leq M-1$) by evaluating the $\ket{f}$ under operator $U'$ instead of $U$.

\subsection{Quantum Pooling Layer}
 The function of pooling layer after the convolutional layer is to reduce the spatial size of the representation so as to reduce the amount of parameters. We adopt average pooling which calculates the average value for each patch on the feature map as pooling layers in our model.  Consider a $2*2$ pixels pooling operation  applied with a stride of 2 pixels. It can be directly realized by ignoring the last qubit  and the $m$-th qubit in quantum context. The input image $\ket{g'}=(g_1,g_2,g_3,g_4,\dots,\dots,g_{M^2})^T$ after this operation can be expressed as the  output image
\begin{align}
  \ket{p}= \left(\sqrt{g_{1}^2+g_{2}^2+g_{M+1}^2+g_{M+2}^2},\sqrt{g_{3}^2+g_{4}^2+g_{M+3}^2+g_{M+4}^2},\dots,\sqrt{g_{M^2-M-1}^2+g_{M^2-M}^2+g_{M^2-1}^2+g_{M}^2}\right)^T.
\end{align}

\subsection{Quantum Fully Connected Layer}
Fully connected layers  complie the data extracted by previous layers to form the final output, it usually appear at the end of convolutional neural networks. We define a parametrized Hamiltonian as the  quantum fully connected layer. This Hamiltonian consists of identity operators $I$ and Pauli operators $\sigma_{z}$,
\begin{align}
\label{qubit_hamiltonian}
\mathcal{H}=h^0I+\sum_{i}h^i\sigma_{z}^i+\sum_{i,j}h^{ij}\sigma_{z}^{i}\sigma_{z}^j+\dots
\end{align}
where $h^0,h^i,h^{ij},\cdots$ are the parameters, and Roman indices $i, j$ denote the qubit on which the operator acts, i.e., $\sigma^i_{z}$ means Pauli matrix $\sigma_{z}$ acting on a  qubit at site $i$. 
We measure the expectation value of the parametrized Hamiltonian  $f(p)=\langle{p}| \mathcal{H} \ket{p}$.
$f(p)$ is the final output of the whole quantum neural network. Then, we add an active function to nonlinearly map $f(p)$ to  $R(f(p))$.  The parameters in Hamiltonian matrix $\mathcal{H}$ are updated by gradient descent method, i.e., are calculated by $\frac{\partial f(p)}{\partial h^i} =\langle{p}| \sigma_{z}^i \ket{p}$ and $\frac{\partial f(p)}{\partial h^{ij}} =\langle{p}| \sigma_{z}^{i}\sigma_{z}^j \ket{p}$.  The parameters in $S$ matrix can be trained through classical backpropagation.

Now, we construct the framework of quantum neural networks. We demonstrate the performance of our method in image processing and  handwritten number recognition in the next section. 

\section{ Numerical Simulations}\label{sec: result}

\begin{figure}
\centering
\includegraphics[width=14cm]{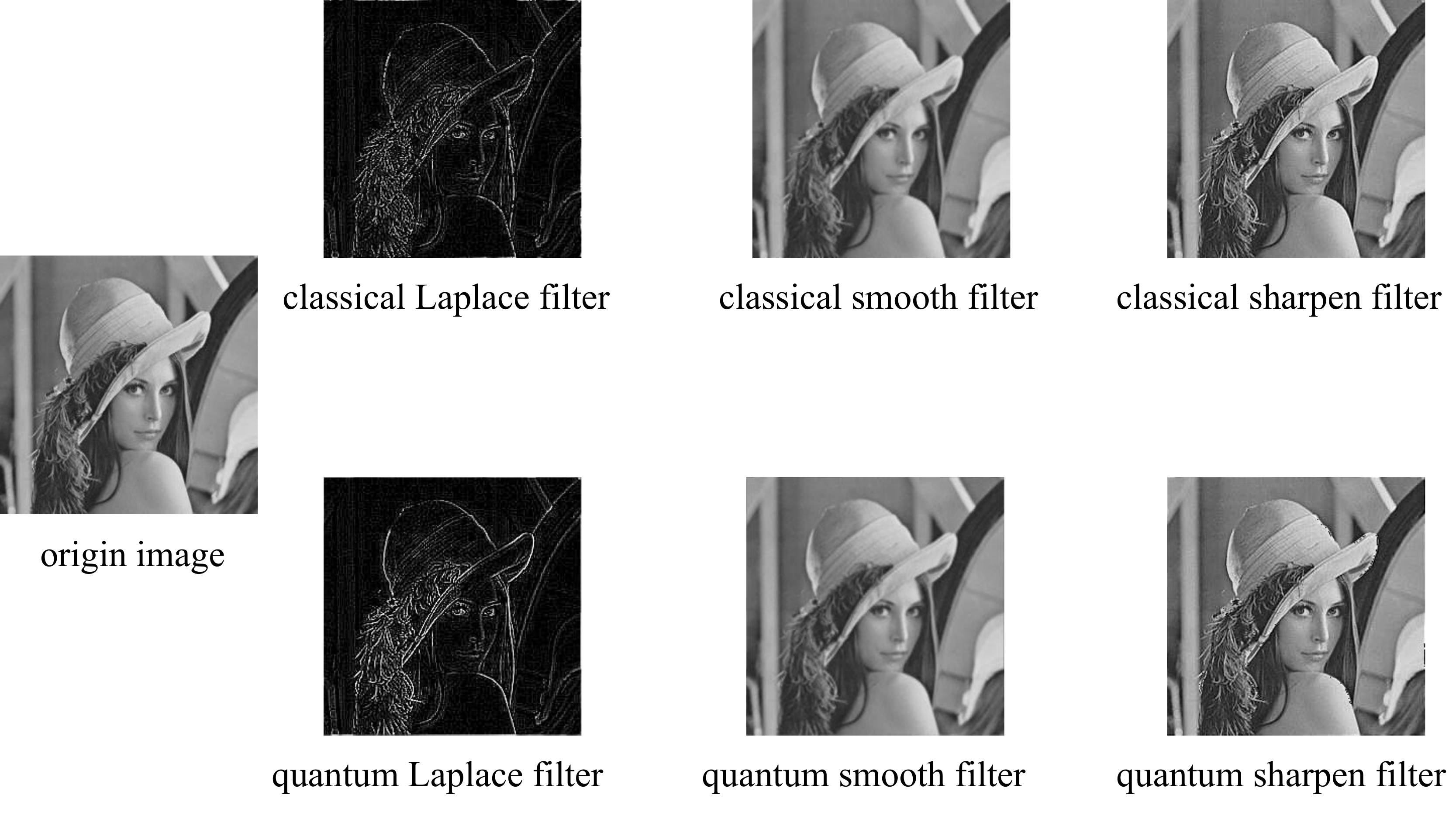}
\caption{Three types of image processing, edge detection, image smoothing and sharpening, are implemented on an image by classical method and quantum method respectively.} \label{exam1}
\end{figure}

\subsection{Image Processing: Edge Detection, Image Smoothing and  Sharpening}
In addition to constructing  QCNN, the quantum convolutional layer can also be used to spatial filtering which is a technique for image processing\cite{yan2016survey,venegas2003storing,le2011flexible,yao2017quantum}, such as image smoothing, sharpening, edge detection and edge enhancement. To show the quantum convolutional layer can handle various image processing tasks, we demonstrate three  types of image processing, edge detection, image smoothing and  sharpening with fixed filter mask $W_{de}$, $W_{sm}$ and $W_{sh}$ respectively
\begin{equation}\label{mask3}
  W_{de}=\left(\begin{array}{ccc}
  -1 & -1 & -1\\
  -1 & 8 & -1\\
  -1 & -1 & -1\\
  \end{array}\right),\\
  W_{sm}=\frac{1}{13}\left(\begin{array}{ccc}
  1 & 1 & 1\\
  1 & 5 & 1\\
  1 & 1 & 1\\
\end{array}\right),\\
  W_{sh}=\frac{1}{16}\left(\begin{array}{ccc}
  -2 & -2 & -2\\
  -2 & 32 & -2\\
  -2 & -2 & -2\\
\end{array}\right).
\end{equation}

In a spatial image processing task, we only need one specific filter mask. Therefore, after performing the above quantum convolutional layer mentioned,  we measure the ancillary register. If we obtain $|0\rangle$, our algorithm succeeds and  the spatial filtering task is completed. The numerical simulation proves that the  output images transformed by a classical and quantum convolutional layer are exactly the same, as shown in Fig.(\ref{exam1}).

\subsection{Handwritten Number Recognition}
 Here we demonstrate a type of image recognition task on a real-world dataset, called MNIST, a handwritten character dataset. In this case, we simulate a complete quantum convolutional neural network model, including a convolutional layer, a pooling layer, and a full-connected layer, as shown in Fig.(\ref{im2}). We consider the two-class image recognition task(recognizing handwritten characters $'1'$ and $'8'$) and ten-class image recognition task(recognizing handwritten characters $'0'$-$'9'$). Meanwhile, considering the noise on NISQ quantum system, we respectively simulate two circumstances that are the quantum gate $Q_k$ is a perfect gate or a gate with certain noise. The noise  is simulated by  randomly acting  a single qubit Pauli gate in $[I, X, Y, Z]$ with a probability of $0.01$ on the quantum circuit after an operation  implemented.
In detail, the handwritten character image of MNIST has $28 \times 28$ pixels. For convenience, we expand $0$ at the edge of the initial image until $32 \times 32$ pixels. Thus, the work register of QCNN consists of 10 qubits and the ancillary register needs 4 qubits. The convolutional layer is characterized by 9 learnable parameters in matrix $W$, that is the same for QCNN and CNN. In QCNN, by abandoning the $4$-th and $9$-th qubit of the work register, we perform the pooling layer on quantum circuit. In CNN, we perform average pooling layer directly. Through measuring the expected values of  different Hamiltonians on the remaining work qubits, we can obtain the measurement values. After  putting them in an activation function, we get the final classification result. In CNN, we perform a two-layer fully connected neural network and an activation function. In the two-classification problem, the QCNN's parametrized Hamiltonian has 37 learnable parameters and the CNN’s fully-connected layer has 256 learnable parameters. The classification result that is close to 0 are classified as handwritten character $'1'$,  and that is close to 1 are classified as handwritten character  $'8'$. In the ten-classification problem, the parametrized Hamiltonian has $10 \times 37$ learnable parameters and the CNN's fully-connected layer has $10 \times 256$ learnable parameters. The result is a 10-dimension vector. The classification results are classified as the index of the max element of the vector. Details of  parameters, accuracy and gate complexity are listed in Table.(\ref{table}).

\begin{figure}[htbp]
  \centering
  \subfigure[]{
  \includegraphics[width=8cm]{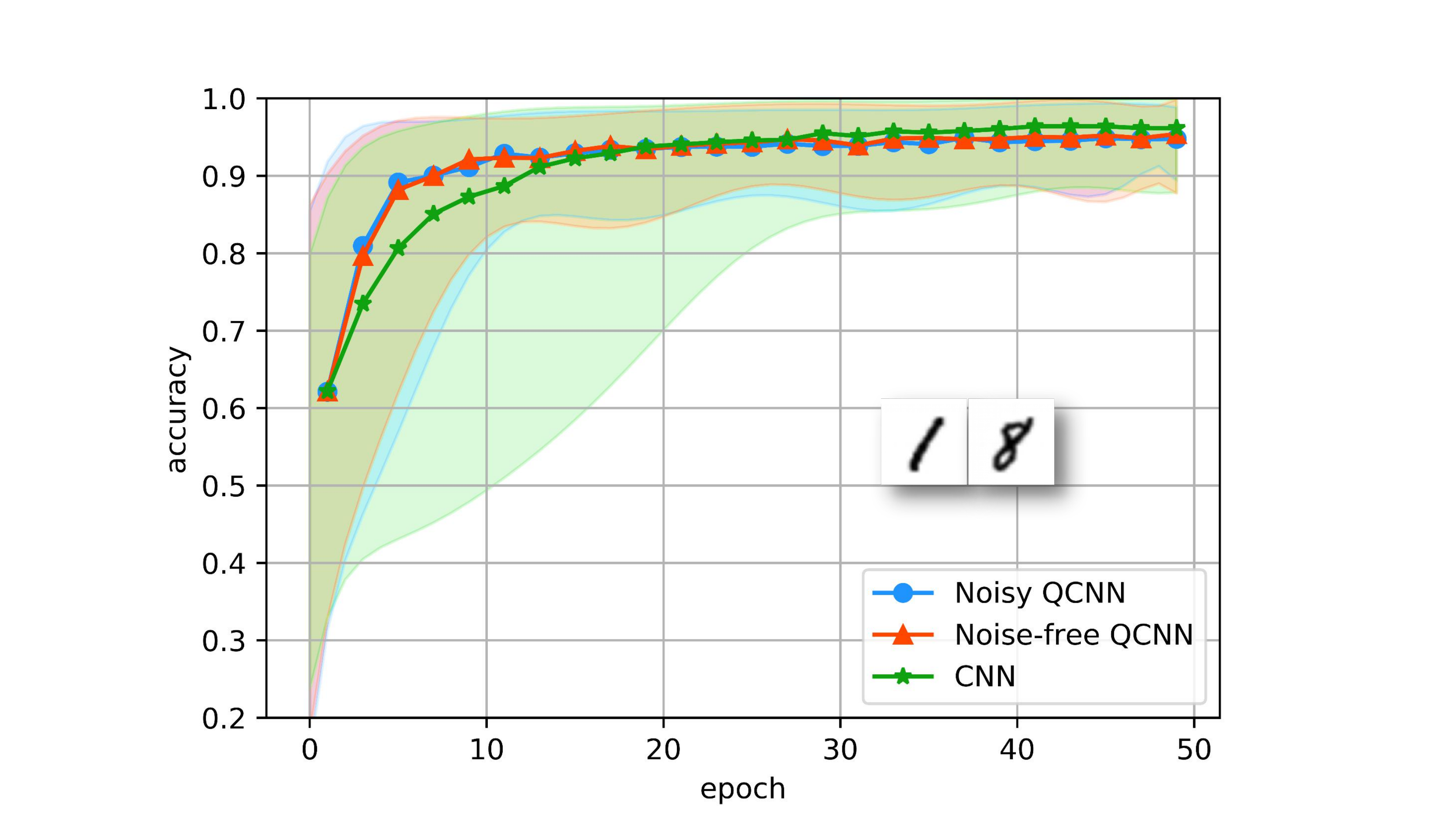}
  }
  \quad
  \subfigure[]{
  \includegraphics[width=8cm]{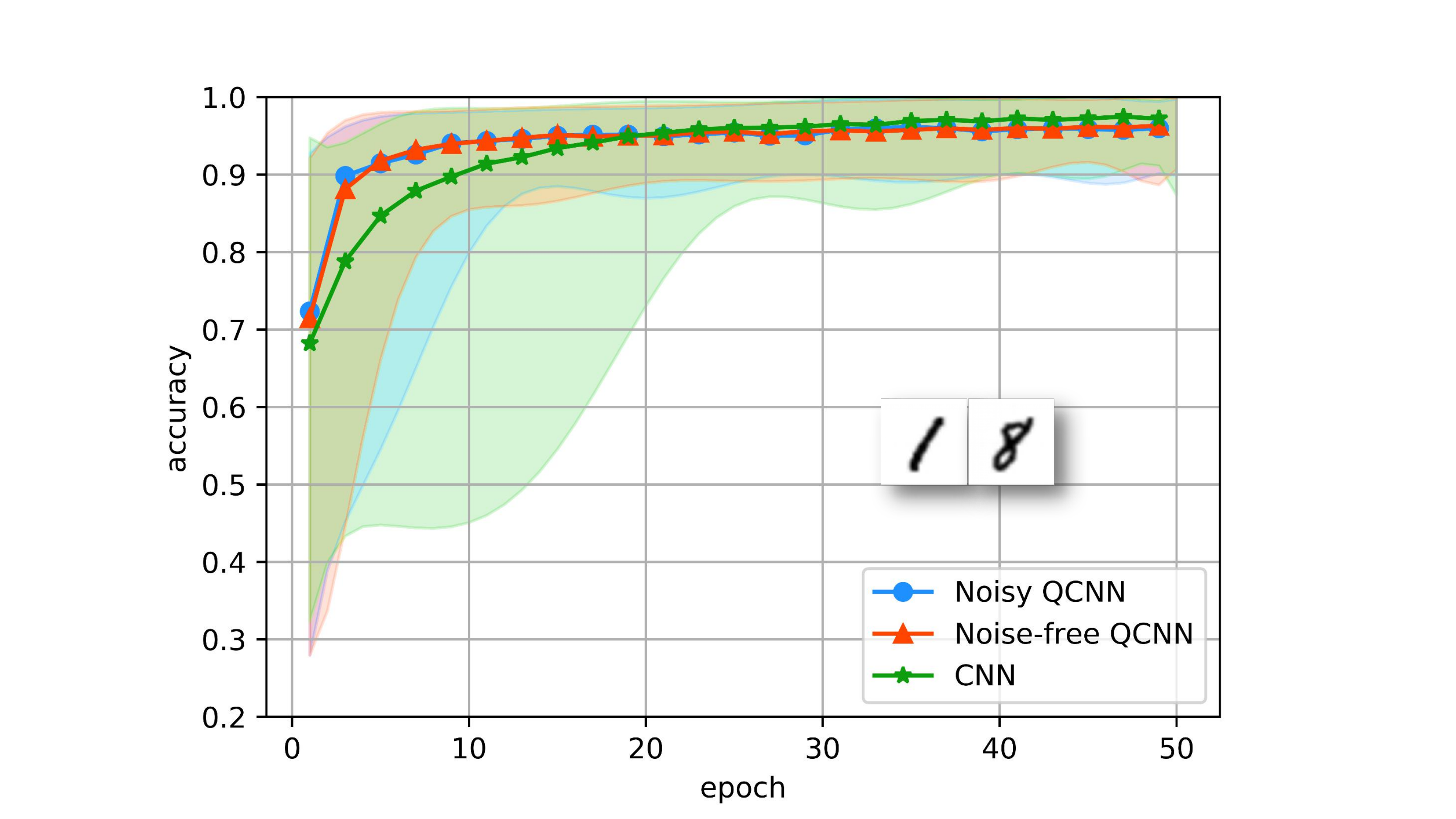}
  }
  \quad
  \subfigure[]{
  \includegraphics[width=8cm]{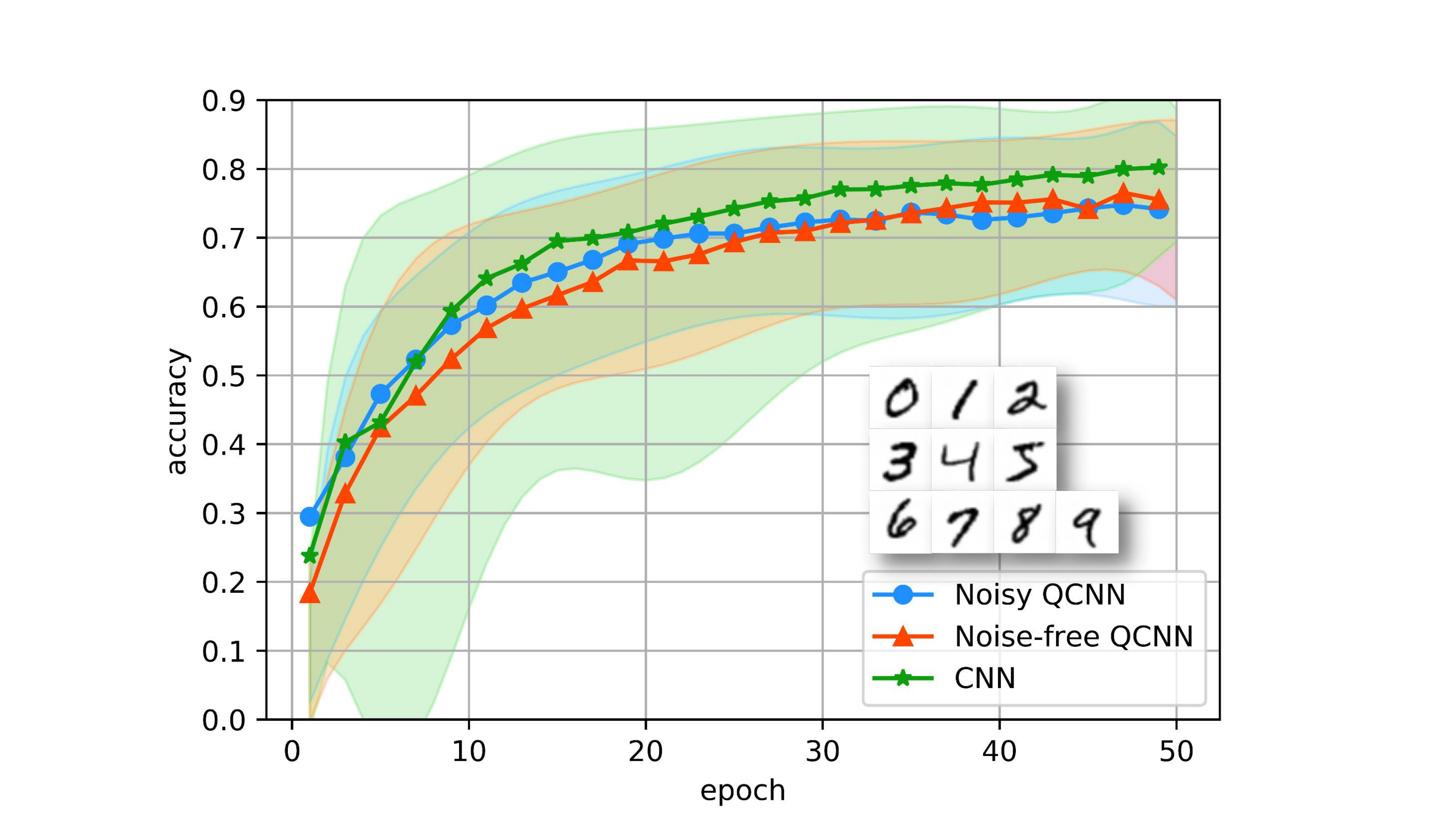}
  }
  \quad
  \subfigure[]{
  \includegraphics[width=8cm]{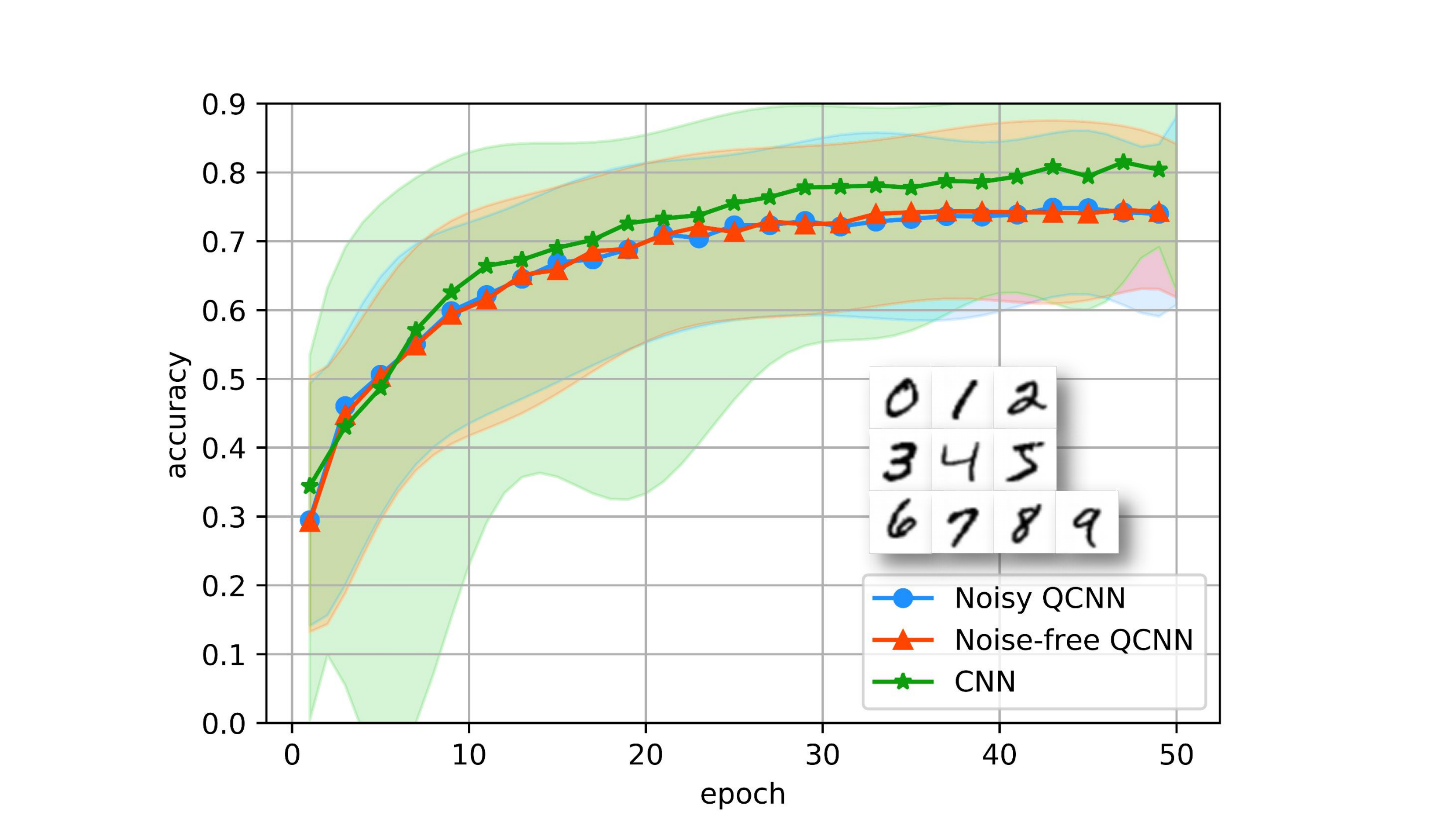}
  }
  \caption{The performance of QCNN based on MNIST. The blue, red, and green curves denote the average accuracy of the noisy QCNN, noise-free QCNN, and CNN, respectively. The shadow areas of the corresponding color denote the accuracy fluctuation range in the 100 times simulation results. The insets are the typical images from MNIST set. (a), (b) are the curves representing the result from the training set and the test set for the 2 class classification problem respectively. (c), (d) show the result from the training set and the test set for the 10 class classification problem respectively.}\label{perf}
  \end{figure}

For the 2 class classification problem, the training set and test set have a total of 5000 images and 2100 images, respectively.  For the 10 class classification problem, the training set and test set have a total of 60000 images and 10000 images, respectively. Because in a training process,  100 images are randomly chosen in one epoch, and 50 epochs in total,  the accuracy of the training set and the test set will fluctuate. So, we repeatedly execute noisy QCNN, noise-free,  and CNN 100 times, under the same construction. In this way, we obtain the average accuracy and the field of accuracy, as shown in Fig.(\ref{perf}).  We can conclude that from the numerical simulation result,  QCNN and CNN provide similar performance. QCNN involves fewer parameters and has a smaller fluctuation range. 

\begin{table}[h] 
\renewcommand\arraystretch{1.5}
\setlength{\abovecaptionskip}{0.cm}
\caption{The important parameters of models}
\begin{tabular}{c|c|cIc|c|c}
\hline
\hline

\multirow{2}{*}{Models} & \multirow{2}{*}{Problems} & \multirow{2}{*}{Data Set} &\multicolumn{3}{c}{Parameters} \\
\cline{4-6} & & &Learnable Parameters & Average Accuracy& Gate Complexity \\

\midrule[1pt]

\multirow{4}{*}{Noisy QCNN}& \multirow{2}{*}{$'1'$ or $'8'$} & training set & \multirow{2}{*}{46} &0.948 &\multirow{8}{*}{$O((log_2M)^6)$}\\
\cline{3-3}\cline{5-5}& & test set &  &0.960& \\

\cline{2-5}
\cline{3-3}\cline{5-5}& \multirow{2}{*}{$'0'\sim'9'$} &training set &\multirow{2}{*}{379} & 0.742& \\
\cline{3-3}\cline{5-5}& & test set &  &0.740& \\

\cline{1-5}

\cline{3-3}\cline{5-5}\multirow{4}{*}{Noise-free QCNN} &\multirow{2}{*}{$'1'$ or $'8'$}& training set & \multirow{2}{*}{46} &0.954 &\\
\cline{3-3}\cline{5-5}& & test set &  &0.963& \\

\cline{2-5}
 \cline{3-3}\cline{5-5}& \multirow{2}{*}{$'0'\sim'9'$} &training set &\multirow{2}{*}{379} &0.756& \\
\cline{3-3}\cline{5-5}& & test set &  &0.743& \\
\cline{1-6}

\cline{3-3}\cline{5-5}\multirow{4}{*}{CNN}& \multirow{2}{*}{$'1'$ or $'8'$} &training set & \multirow{2}{*}{265} &0.962&\multirow{4}{*}{$O(M^2)$} \\

\cline{3-3}\cline{5-5}& & test set &  &0.972 & \\

\cline{2-5}
\cline{3-3}\cline{5-5}& \multirow{2}{*}{$'0'\sim'9'$ } &training set &\multirow{2}{*}{2569} & 0.802 & \\
\cline{3-3}\cline{5-5}& & test set &  &0.804 & \\

\hline
\hline

\end{tabular}\label{table}
\end{table}

\section{Algorithm complexity }
We  analyze the computing resources in gate complexity and qubit consumption. (1) Gate complexity. At the convolutional layer stage, we could  prepare an initial state in $O(poly(log_{2}(M^2))$ steps. In the case of preparing a particular input  $\ket{f}$, we employ the amplitude encoding method in Ref.\cite{long2001efficient,grover2002creating,soklakov2006efficient}. It was shown that if the amplitude $c_k$ and $P_k=\sum_k |c_k|^2 $ can be efficiently calculated by a classical algorithm, constructing the $log_{2}(M^2)$-qubit $X$ state  takes $O(poly(log_{2}(M^2))$ steps. Alternatively, we can resort to quantum random access memory\cite{giovannetti2008quantum,giovannetti2008architectures,arunachalam2015robustness}. Quantum Random Access Memory (qRAM) is an efficient method to do state preparation, whose complexity is $O(log_{2}(M^2))$ after the quantum memory cell established. Moreover, the controlled operations $ Q_{k} $  can be decomposed into $ O((log_{2}M)^6) $ basic gates (see details in Appendix A). In summary, our algorithm uses  $ O((log_{2}M)^6) $ basic steps to realize the filter progress in the convolutional layer.
For CNN, the complexity of implementing a classical convolutional layer is $ O(M^2) $. Thus, this algorithm achieves an exponential speedup over classical algorithms in gate complexity. The measurement complexity in fully connected layers is  $ O(e)$, where $e$ is the number of parameters in the Hamiltonian.\\
 (2) Memory consumption. The ancillary qubits in the whole algorithm  are $ O(log_{2}(m^2)) $, where $m$ is the dimension of the filter mask, and the work qubits are $ O(log_{2}(M^2))$. Thus, the total qubits resource needed is $ O(log_{2}(m^2)+O(log_{2}(M^2) $. 

\section{Conclusion}\label{sec:Conclu}
In summary, we desighed a quantum Neural Network which provides exponential speed-ups over their classical counterparts in gate complexity. With  fewer parameters, our model achieves similar performance compared with classical algorithm in handwritten number recognition tasks. Therefore, this algorithm has significant advantages over the classical algorithms for large data. We present two interesting and practical applications, image processing and handwritten number recognition, to demonstrate the validity of our method. We give the mapping relations between a specific classical  convolutional kernel to a quantum circuit, that provides a  bridge between QCNN to CNN. It is a  general algorithm and can be implemented on any programmable quantum computer, such as superconducting, trapped ions, and photonic quantum computer. In the big data era, this algorithm has great potential to outperform its classical counterpart, and works as an efficient solution.

\bigskip\noindent
\textbf{Acknowledgements}
\begin{acknowledgments}
  We thank X. Yao,  X. Peng for inspiration and fruitful  discussions. 
  This research was supported by National Basic Research Program of China.  S.W. acknowledge  the China Postdoctoral Science Foundation 2020M670172 and the National Natural Science Foundation of China under Grants No. 12005015. We gratefully acknowledge support from the National Natural Science Foundation of China under Grants No. 11974205, and No. 11774197. The National Key Research and  Development Program of China (2017YFA0303700); The Key Research and  Development Program of Guangdong province (2018B030325002); Beijing Advanced Innovation Center for Future Chip (ICFC).
\end{acknowledgments}


\appendix

\section*{Appendix A: Adjusted operator $U'$ can provide enough information to remap the output imagine.}
$ Proof.$-
The different elements of image matrix after implementing operator $U'$ compared with $U$ are in the edges of image matrix. We prove that the evolution under operator $U'$ can provide enough information to remap the output image.
The different elements between $U'$ and $U$ are included in 

\begin{equation}
  U'_{k,n}\neq U_{k,n} \begin{cases}(1\leq k \leq M; 1\leq n \leq 2M, M^2-M\leq n \leq M^2)\\
(M^2-M\leq k \leq M^2; 1\leq n \leq M,M^2-3M\leq n \leq M^2-M)\\
(k=sM+1; n=1+(s-1)M,2+(s-1)M,sM,sM+1,sM+2,(s+1)M,(s+1)M+1,(s+1)M+2,(s+2)M)\\
(k=(s+1)M;n=1,sM-1,sM,sM+1,(s+1)M-2,(s+1)M-1,(s+1)M+1,(s+2)M-2,(s+2)M-1)
\end{cases}\label{eq:U}
\end{equation}
where $ 1\leq s \leq M-2$.

After performing $U'$ and $U$ on quantum state $\ket{f}$ respectively, the difference exits in the elements $ \ket{g'_{k}}\neq \ket{g_{k}}(k=1,2,\cdots,M,sM+1,(s+1)M,M^2-M+1,\cdots,M^2)$, where $1\leq s \leq M-2$. 
Since $\ket{g'}$ can be remapped to $G'$, $U'$ will give the output image $G'=(G'_{i,j})_{M \times M}$.  The elements in $U'$  which is different from $U$ only affect the pixel  $ i, j \not\in {2, \cdots, M-1 }$. Thus, only and if only  $ i, j\not\in {2, \cdots, M-1 }$, the matrix elements satisfy $G_{i,j} \neq G'_{i,j}$. Namely, the output imagine $G'_{i,j}=G_{i,j}$($2\leq i, j \leq M-1$).

\section*{Appendix B: Decomposing operator Q into basic gates}
Considering the nine operators $Q_1$, $Q_2$,$\cdots$, $Q_9$ consist of  filter operator $U'$.
$Q_k$ is the tensor product of two of the following three operators

\begin{equation}
  E_{1 }=\left(
  \begin{array}{cccccc}  
   &  & &   &  1     \\
  1 &     &    &  &   \\
  \ddots & \ddots & \ddots &  &\\
  & &  1 & & \\
   & &   & 1 &   \\
  \end{array}\right)_{M\times M}\\
E_{2 }=\left(
  \begin{array}{cccccc}  
  1  &   &   &  &      \\
   & 1 &      &  &   \\
   &   \ddots & \ddots &\ddots & \\
   & &   &1 & \\
  & &  &  & 1     \\
  \end{array}\right)_{M\times M}\\
  E_{3}=\left(
  \begin{array}{cccccc}  
   &  1 & &     &     \\
   &  & 1  &      &   \\
   &  &\ddots \ddots & \ddots & \\
   & &   & & 1\\
   1  & &   &   &   \\
  \end{array}\right)_{M\times M}.
  \end{equation}

\begin{figure}
\centering
\includegraphics[width=0.8\textwidth]{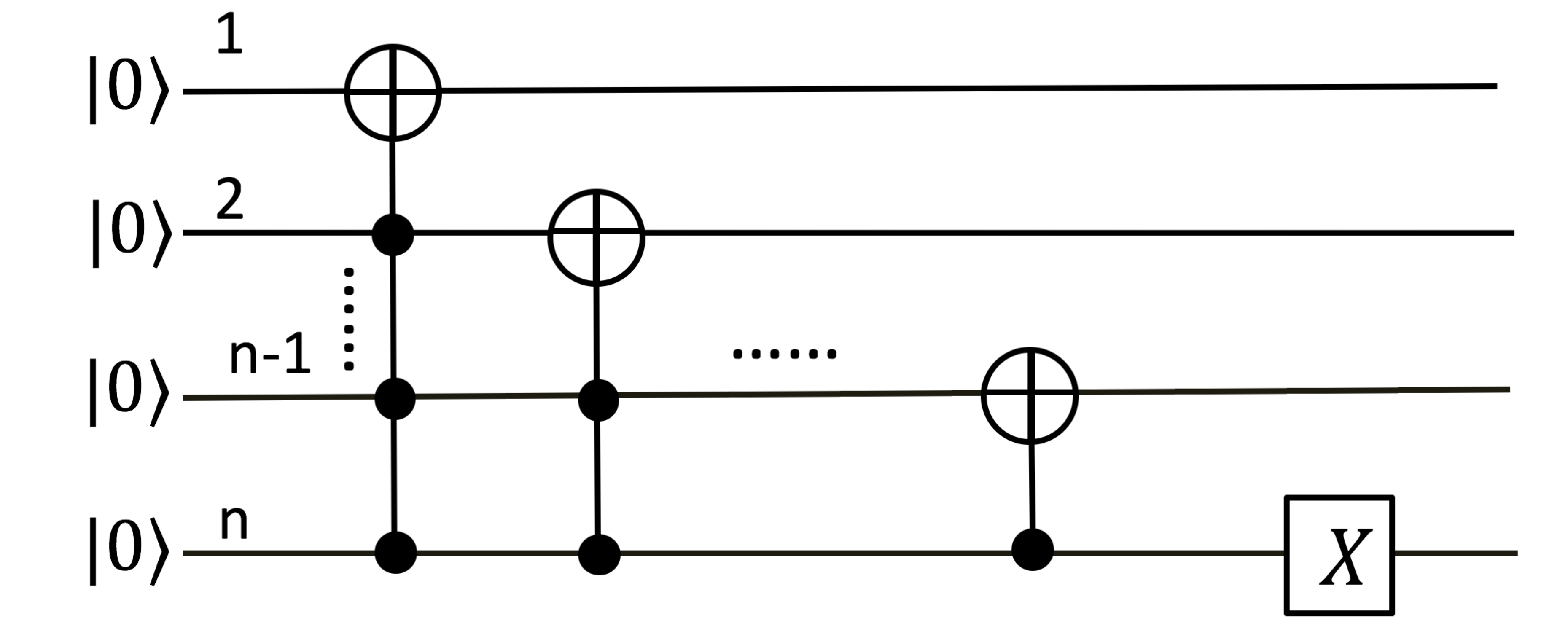}
\caption{Decomposition of operator $E_1$ in the form of basic gates.} \label{ap1}
\end{figure}

$E_{2}$ is a $M\times M$ identity matrix  not need to be further decomposed. For convenient, consider a $n$-qubits operator $E_1$ with dimension $M\times M$, where $n=log_2(M^2)$. It can be expressed by the combination of $O(n^3)$ CNOT gates and Pauli $X$ gates as shown in Fig.{\ref{ap1}}. Consequently, $E_3$ can be decomposed into the inverse of combinations of basic gate as shown in Fig.{\ref{ap1}}, because of the fact $E_3=E_1^{\dagger}$. 
Thus, $Q_k$ can be implemented by no more than $O(n^6)$ basic gates. Totally, the controlled $Q_k$ operation can be implemented by  no more than $O(n^{6})=O((log_2M)^{6})$(ignoring constant number).

\bibliographystyle{unsrt}

\bibliography{QCNN}

\end{document}